\documentclass[epj]{svjour}
\usepackage{graphics}
\begin{document}
\title{A fundamental threat to quantum cryptography:
gravitational attacks}
\author{R. Plaga\inst{1}
\thanks{\emph{E-mail:} rainer.plaga@bsi.bund.de}%
}                     
%
%
\institute{Federal Office for Information Security (BSI), 53175 Bonn, Germany}
\date{Received: date / Revised version: date}
%
\abstract{
An attack on the ``Bennett-Brassard 84''(BB84) quantum key-exchange protocol 
in which Eve exploits the action of gravitation to infer information about the
quantum-mechanical state of the qubit exchanged between Alice and 
Bob, is described. It is demonstrated that the known laws of physics
do not allow to describe the attack. Without making assumptions that are not based 
on broad consensus, the laws of quantum gravity, unknown up to now, would be needed
even for an approximate treatment. Therefore, it is currently not possible to predict with any
confidence if information gained in this
attack will allow to break BB84. Contrary to previous belief, a proof of the perfect 
security of BB84 cannot be based on the assumption that the known laws of physics
are strictly correct, yet. 
\\
A speculative 
parameterization that characterizes the time-evolution operator of quantum gravity
for the gravitational attack is presented. It allows to evaluate
the results of gravitational attacks on BB84 quantitatively. 
It is proposed to perform state-of-the-art gravitational attacks,
both for a complete security assurance of BB84 and as an unconventional search
for experimental effects of quantum gravity. 
\PACS{
      {03.67.Dd}{Quantum cryptography}   \and
      {04.60.-m}{Quantum gravity}        \and
      {03.65.Ta}{Foundations of quantum mechanics, measurement theory}
     } 
} 
\maketitle

\section{Introduction}
\label{abstract}
\newtheorem{thm1}{}
\newtheorem{thm2}{}
Quantum key-distribution (QKD) protocols, often collectively called
``quantum cryptography'', exploit the principles of quantum mechanics
to enable the secure distribution of information\cite{qc}.
It is a common belief that the perfect secrecy 
of keys exchanged by such protocols is 
guaranteed if the ``known laws of physics''\footnote{Defined
here as an expression that was derived from a consistent
mathematical framework (a ``theory of physics'')
and has been confirmed 
by repeated scientific experiments.} 
are assumed to be strictly correct\cite{lo_talk,christandl04}.
This would be a major advantage of quantum
cryptography because an analogous
security guarantee for classical cryptography - based on the
correctness of proven, or at least highly
plausible, mathematical theorems\footnote{The analogues to laws
of physics.} - is not possible, yet\cite{goldreich}.
\\
Section \ref{attackd}
presents a novel attack procedure against the first and best known
QKD protocol, the ``Bennett-Brassard 84'' (BB84) protocol\cite{qc},
in which the attacker exploits the action of gravity.
I demonstrate in section \ref{qgrav} that
this attack cannot be modelled - not even to any
approximation - on the basis of the known laws of physics 
without making assumptions that are not based on broad consensus.
\\
Even though its security proof
is shown to be incomplete, BB84 retains its great value because
it rests on completely different foundations than its classical
counterparts. However,
for a complete security assurance one needs to attack the protocol experimentally.
In section \ref{nonlinear} I propose a
framework in which the results of gravitational attacks on BB84
can be evaluated quantitatively. In this framework Eve breaks BB84
via gravitationally cloning a qubit,
section \ref{special} studies if this indirectly violates 
special relativity.
Section \ref{concl} concludes.

\section{The ``gravitational-attack'' protocol}
\label{attackd}
In the BB84 protocol the honest party (``Alice'') encodes a bit
of the key to be distributed by
preparing
a qubit ``Q'' either in one of the four quantum-mechanical states 
$|\Psi>$ = $|0>$, $|\Psi>$ = $|1>$, $|\Psi>$ =$|+>$ = ${1 \over \sqrt{2}}$($|0>$ + $|1>$)  or 
$|\Psi>$ = $|->$ = ${1 \over \sqrt{2}}$($|0>$ -- $|1>$). 
She then sends Q to its designated 
receiver (``Bob'').
Rigorous proofs of the
security of BB84\cite{lo,mayers} are based on the
assumption that the laws of quantum
physics are correct.
However, these proofs ignore gravitation.
Implicitly they assume that attackers
only employ the resources of quantum physics in flat
space time.
However, it seems overly optimistic to 
``require'' eavesdroppers to avoid the profound
difficulties that still beset
any attempt to definitely answer the question: ``What
gravitational field corresponds to a given quantum state?''
\\
In a ``gravitational attack'' the eavesdropper (malicious ``Eve'')
employs a beam splitter to evolve Q into a state:
\begin{equation}
|\Psi> = 1/\sqrt{2}(|\Psi(x_1)> + |\Psi(x_2)>)
\end{equation}
consisting of two spatially separated components at 
the spatial positions x$_1$ and
x$_2$, respectively. She then measures the state of $|\Psi(x_1)>$
in one of the bases employed in the BB84 protocol ($|0>,|1>$) and
the state of $|\Psi(x_2)>$ in the other base ($|+>,|->$). Depending
on which of the four possible measurement results 
$|s>$ (with s=0, s=1, s=+ or s=--) is obtained, a
macroscopic test mass M, initially
at the spatial position x(1), is automatically moved to 
(or left at) one of four separated spatial positions
x(0),x(1),x(+),x(--). Immediately thereafter Eve 
experimentally determines the gravitational field surrounding these four positions,
e.g. with the help of a Cavendish setup. 
Can we derive a definite prediction for
the results of Eve's field-strength determination, 
based on the known laws of physics?
\section{The attack cannot be described by the known laws of physics - not
even approximately}
\label{qgrav}
%
The only ``known law of physics'' that describe
gravitation are classical: they derive from Einstein's theory of general
relativity\cite {einstein}. The only theoretical ansatz to describe
the attack
that keeps both general relativity and quantum physics
unchanged is ``semiclassical gravity".
It proposes that
the source of the gravitational field is the quantum
expectation value of the energy tensor of matter\cite{qg}:
\begin{equation}
G_{\mu\nu} = 8 \pi {\rm G/c^2} <M|T_{\mu\nu}|M>
\label{sg}
\end{equation}
here $G_{\mu\nu}$ is the classical Einstein tensor,
G is Newton's constant of gravitation, c the 
speed of light, $T_{\mu\nu}$ the
stress-energy tensor and $|M>$ the quantum mechanical state
of the gravitating body.
However, this expression cannot
be considered to be even an established approximation to a law of nature.
If there is no wave-function collapse and
standard quantum physics allows a complete description of nature, i.e.
if the ``many-worlds'' interpretation (MWI) of quantum mechanics\cite{everett,zeh70} is correct,
eq.(\ref{sg}) predicts a nonlinear coupling of quantum-mechanical state components\cite{page81}
(see section \ref{semi} for further explanation).
Page and Geilker\cite{page81} presented experimental data that
rule out such a coupling at the strength expected from
eq.(\ref{sg}) with a high confidence level. 
The nonlinear coupling does not vanish in the low-energy or
weak-field limit of eq.(\ref{sg}). Within the MWI 
eq.(\ref{sg}) is wrong even to the approximation that general relativity describes
gravitation. Page and Geilker drew the conclusion that there must be 
as yet unknown laws
of physics, beyond semiclassical general relativity, that describes their experiment.
\\
The quantum-information community is currently not in a state of
agreement whether the MWI is correct, but
some of its eminent members 
advocated this idea\cite{deutschb,preskill229,dewitt}, and many specialists
at least admit the principal possibility that it might be correct\cite{privQICL}.
The assumption that this interpretation is wrong clearly would be not
based on a broad consensus and can therefore serve  
neither as a basis for a prediction of the outcome of the attack nor
for any sound security proof. 
\\
More complicated schemes to couple a classical gravitational
field to the quantized matter field might be possible, but
have not been proposed, yet, to my knowledge. 
It is generally considered much more likely that
general relativity will turn out to be the limit of $\hbar$ $\rightarrow$ 0
of a theory of ``quantum gravity'', that remains to be
discovered.
However, due to various technical
and conceptual difficulties, 
all candidate theories of quantum
gravity\cite{qg} still fall far short
of a reliable basis for deriving ``known law of physics''.
Moreover none of the nonperturbative approaches to this problem
have obtained a definite classical limit, yet.
Thus, even if one of them were a correct theory
of physics, it would not be possible 
to derive predictions for the attack, yet. 
In particular there is
no basis on which certain properties of the known laws
of quantum mechanics, like e.g. its linearity, can be
assumed to hold for quantum gravity.
\\
Summarizing, in the possible case that the
``many-worlds interpretation'' is correct,
even for a qualitative prediction of the result
of Eve's measurement a theory of quantum gravity is needed.
All proofs of the security of BB84 remain incomplete
because a definite theoretical basis 
to address the question ``What information 
can Eve extract from the exchanged qubit in a
``gravitational attack''?'' does not exist presently. 
With other words:
our failure to understand quantum gravity
prevents a basic condition for any security assurance 
to be met for QKD, yet: the target
of evaluation must be thoroughly understood, also 
when being under attack.

\section{An ``insecure-BB84'' scenario: nonlinear 
quantum gravity}
\label{nonlinear}
To illustrate how laws of quantum gravity could
render BB84 brittle,
to motivate experimental attacks on this protocol,
and to supply a framework for their analysis, 
I characterize a speculative time-evolution operator of quantum gravity
for the gravitational attack 
in section $\ref{gen}$. This is not meant
as a serious proposal for a theory of quantum gravity, but
merely as parameterization to allow a quantitative 
analysis of ``gravitational attacks''.
Nonlinearity was chosen only as an example. There might be
other characteristics of quantum gravity that render
QKD vulnerable.
\\
Section \ref{semi} reviews semiclassical gravity, already
discussed in section \ref{qgrav}, more formally. Under
the assumptions discussed in section \ref{qgrav}, this theory
predicts a strongly nonlinear evolution during and after the ``gravitational attack''.
As the opposite extreme section \ref{lin} presents a completely
linear form of the time-evolution operator of quantum gravity.
In section \ref{gen} I propose
a ``general'' time-evolution operator that 
interpolates between these two cases.
\\
For illustration let us 
always assume below that initially Alice prepares the exchanged qubit 
in the BB84 protocol in the state
$|\Psi>$=$|1>$.
\\
\subsection{Semiclassical gravity}
\label{semi}
Let us first assume that the gravitational field remains a classical field
even at the fundamental level, i.e. that eq.(\ref{sg}) is a law of physics.
As in section \ref{qgrav} we assume the MWI.
The initial ``state''\footnote{This is not
a quantum mechanical state in the usual sense but a juxtaposition
of quantum-mechanical and classical fields.} of the 
system of qubit Q and test mass M is given as:
\begin{equation}
|\phi_{\rm semiclassical \ gravity}>(t=0)> = |1> \otimes |M(1)> G_{\mu\nu}(1)
\end{equation}
Here and in the following $|s>$ denotes the state of a qubit exchanged in BB84, 
and $|M(s)>$ the one of the macroscopic test mass.
G$_{\mu\nu}$(s) is the classical Einstein tensor, that characterizes
the structure of space time with an isolated macroscopic test mass M(s)
at the spatial
position x(s). s denotes the state of the exchanged
qubit Q according to the attack protocol (see section \ref{attackd}).
According to eq.(\ref{sg}):
\begin{equation}
G{_{\mu\nu}}(s) = 8 \pi {\rm G/c^2} <M(s)|T_{\mu\nu}|M(s)>
\label{sg_lin}
\end{equation}
The exchanged qubit is
neglected in this expression because of its usually very small mass energy. 
The quantum-mechanical state after the gravitational attack 
(section \ref{attackd}) is given as:
\begin{eqnarray}
|\phi_{QM}>(t=t_f) =
{1 \over \sqrt{2}} |1> \otimes |M(1)>  +
\nonumber
\\
1/2 (|+> \otimes |M(+)> + |-> \otimes |M(-)>) 
\label{QM}
\end{eqnarray}
Including the gravitational field one obtains:
\begin{eqnarray}
|\phi_{\rm semiclassical \ gravity}>(t=t_f) = V_{\rm sg} |\phi>(t=0) = 
\nonumber
\\
{1 \over \sqrt{2}} |1> \otimes |M(1)>  +
\nonumber
\\
1/2 (|+> \otimes |M(+)> + |-> \otimes |M(-)>) 
G_{\mu\nu}(\phi_{QM})
\label{attack_sg}
\end{eqnarray}
The classical Einstein tensor
$G_{\mu\nu}(\phi_{QM})$ 
characterizes 
the gravitational field exerted by all three mass components M(1),M(+) and M(--).
It can be evaluated by inserting eq.(\ref{QM}) into eq.(\ref{sg}).
Because cross terms rapidly vanish due to decoherence,
the source of this gravitational-field are the expectation values
of the energy tensor of the three masses and one obtains:
\begin{eqnarray}
G_{\mu\nu}(\phi_{QM}) = 
1/2 G_{\mu\nu}(1)
+1/4 G_{\mu\nu}(+) + 1/4 G_{\mu\nu}(-)
\label{sg_nonlin}
\end{eqnarray} 
\\
The further evolution of this state is strongly nonlinear due to the
gravitational coupling, i.e. V$_{\rm sg}$ cannot be a linear
operator, but must be some different nonlinear operator of quantum gravity.

\subsection{Linear quantum gravity}
\label{lin}
Alternatively the hypothetical gravitational quantum field
could obey an equation of motion, that is
precisely linear - like all other
known quantum fields do.
The initial state of the setup before the attack is then written as:
\begin{equation}
|\phi>(t=0) = |1> \otimes |M(1)> \otimes |G_{\mu\nu}(1)>
\label{lin_ini}
\end{equation}
$|G_{\mu\nu}(s)>$ symbolizes a hypothetical ``quantum state 
of the gravitational
field'' that is characterized by a space-time structure
described by the classical Einstein tensor G$_{\mu\nu}(s)$ 
(eq.(\ref{sg_lin})) that
describes space time for an isolated
test mass M(s) at spatial position x(s). 
\\
U$_{\rm lqg}$ be a linear unitary operator.
The final state at time
t$_f$ after the attack
described in section \ref{attackd} is then given as:
\begin{eqnarray}
|\phi_{\rm linear \ quantum \ gravity}>(t_f)= U_{\rm lqg} |\phi>(t=0) = 
\nonumber
\\
{1 \over \sqrt{2}} |1> 
\otimes |M(1)> \otimes |G_{\mu\nu}(1)> +
\nonumber
\\
1/2 (|+> \otimes |M(+)> \otimes |G_{\mu\nu}(+)> + 
\nonumber
\\
|-> 
\otimes |M(-)> \otimes |G_{\mu\nu}(-)>)
\label{attack_qg}
\end{eqnarray}
The further evolution of this state will be
linear.
\subsection{General quantum gravity}
\label{gen}
If the MWI interpretation is correct,
it is experimentally excluded 
that eq.(\ref{attack_sg}), that is initially relatively well 
motivated theoretically\footnote{Since it only combines ``known laws of physics''.}, 
is correct (section \ref{qgrav}).
On the other hand,
the assumption of strictly linear U$_{lqg}$
in eq.(\ref{attack_qg}), that is in agreement with all
available data, 
lacks any theoretical basis.
It has indeed been recently speculated that quantum gravity is
nonlinear\cite{wang}. 
\\
Nonlinear effects 
might not be negligible 
even if they occur only near
the Planck energy scale M$_{\rm Planck}$.
Phenomenological effects
at familiar energies would are
typically suppressed by a factor s=(m$_p$/M$_{\rm Planck})^2$,
where m$_p$ is the proton mass. Recently
string theories with large extra dimensions,
in which the Planck scale
M$_{\rm Planck}$ might be as small as 1 TeV,
have been developed\cite{cavaglia}. 
The suppression factor s might thus be of respectable magnitude for
energies commonly encountered in the laboratory.
\\
Clearly a plausible general phenomenological ansatz 
for time evolution in quantum gravity must allow for the possibility
of nonlinearity. 
Let us assume the initial state of eq.(\ref{lin_ini}).
As the final state I propose a combination of 
of the linear eq.(\ref{attack_qg}) and the semiclassical
eq.(\ref{attack_sg}):
\begin{eqnarray}
|\phi_{\rm general \ quantum \ gravity}>(t=t_f) = 
\nonumber
\\
V_{\rm gqg}|\phi>(t=0) = 
\nonumber
\\
{1 \over \sqrt{2}} |1> \otimes 
|M(1)> \otimes |G^{\rm general}_{\mu\nu}(1)> +
\nonumber
\\
1/2 (|+> \otimes |M(+)> \otimes |G^{\rm general}_{\mu\nu}(+)> + 
\nonumber
\\
|-> \otimes 
|M(-)> \otimes |G^{\rm general}_{\mu\nu}(-)>)
\label{attack_fin}
\end{eqnarray}
with the classical Einstein tensor:
\begin{eqnarray}
G^{\rm general}_{\mu\nu}(s) = G_{\mu\nu}(s) +
b e^{- \lambda \Delta t} 
G_{\mu\nu}(\phi_{QM})
\label{sg_gen}
\end{eqnarray} 
$G_{\mu\nu}(s)$ is given by eq.(\ref{sg_lin}) and 
$G_{\mu\nu}(\phi_{QM})$ by
eq.(\ref{sg_nonlin}).
b and $\lambda$ are both purely phenomenological constants.
b $<$ 1 is the amplitude of a ``nonlinear component'' and
1/$\lambda$ a time scale
on which the nonlinear component of the gravitational field is assumed to decay
spontaneously after it first appears due to some mass movement.
$\Delta$t=t-t$_f$ is the time since moving the masses to their respective
spatial positions, i.e. after the end of the attack.
The evolution of this state is nonlinear due to a 
gravitational coupling with an amplitude b e$^{- \lambda \Delta t}$, i.e.
V$_{\rm gqg}$ can be a linear operator only to some approximation.
\\
The gravitational field described by the 
second term 
in eq.(\ref{sg_gen}) is determined
by all three components of the test-mass state even after Eve measured Q. 
From eq.(\ref{sg_nonlin}) one reads that the component with the largest
tensor amplitude in the second term (in our example $|1>$)
corresponds to the state in which Alice prepared
the qubit. Via experimentally determining 
the exact structure of the second term, 
Eve can thus infer
the state of the exchanged qubit. 
\\
She is then able to construct a
clone of the exchanged qubit and sends it
to Bob. BB84 is now broken, because Eve disposes of
the same resources as Bob who cannot detect her
eavesdropping.
In the scenario Eve's attack
exploits an EmSec vulnerability:
Q can be cloned
due to the uncontrolled emission of static gravitational
fields.
\\
I constructed the framework of section \ref{gen}
wearing the hat of 
a security specialist, not the one of a research
scientist. The latter would tend to make assumptions
that allow a consistent
understanding of the attack:
either the standard interpretation 
of quantum mechanics or
a quantum theory of gravity that is strictly
linear. The former tries to 
endanger the security of BB84
with ideas that
are reasonably plausible and 
are clearly not in conflict with the known laws of 
physics:
the ``many-worlds interpretation''
of quantum mechanics and nonlinear quantum
gravity.
\\
I propose to perform the attack described in section \ref{attackd} 
as sensitive and on a time scale as short as possible with state-of-the-art
equipment.
The results of such an experiment can be used to set an upper limit on
b and
$\lambda$ in eq.(\ref{attack_fin}), respectively.
\\
For $\lambda$=0
the experimental results of Page and Geilker\cite{page81} limit b
to be smaller than about 0.1. However, an attacker who exploits state-of-the-art
methods could explore magnitudes of b several orders
of magnitude smaller. 
\\
Sensitive limits on b and $\lambda$ would be an empirical assurance that BB84 is secure
against gravitational attacks.
Our trust in BB84 could then be
analogous to the one conferred to classical cryptographic procedures
by dedicated but unsuccessful attempts of 
highly qualified personnel to break them.
In both cases there is no guarantee that an attacker might not find some
creative, unexpected way to break the protocol.

\section{A successful attack does not need to violate special relativity}
\label{special}
The illustrative successful attack option described in 
section \ref{gen} involved the cloning
of a quantum state. The ``no-cloning'' theorem forbids
this, but its proof\cite{qc} assumes the linearity of temporal
evolution that is guaranteed by the laws of conventional quantum mechanics 
but might not hold in quantum gravity.
\\
More generally it was argued that any successful cloning of quantum states would necessarily
enable superluminal signalling\cite{gisin}. If that were true, a successful attack
would appear to be ruled out under the usually stated assumptions
for quantum cryptography, because superluminal
signalling contradicts the ``no-signalling theorem''
a known law of physics that can be derived from 
special relativity.
However, Kent\cite{kent_mixed} has recently argued that a procedure that allows 
the cloning of pure, localized states, but not the cloning of subsystems
of ``non-local'' mixed states, avoids the argument above. Moreover 
Polchinski\cite{polchinski} has shown that if the MWI is correct,
universal cloning leads to the possibility of communication between macroscopic
components of the total wavefunction, rather than superluminal signalling.
Such an ``Everett phone'' would neither be in obvious contradiction with
any known law of physics nor would it lead to counterintuitive
effects if the time scale 1/$\lambda$ in eq.(\ref{sg_gen}) is sufficiently short. 

\section{Summary and outlook}
\label{concl}
It is well known that the security of quantum cryptography 
could be compromised if the laws of quantum mechanics are
not strictly correct\footnote{This realization has
led to recent proposals for QKD protocols that are claimed to remain
secure even if the laws of quantum mechanics are not
strictly correct\cite{barrett}.}, e.g. if the 
usual quantum-mechanical operator ``U'', describing temporal evolution,
would contain a small nonlinear term. However, the assumption
of its strict linearity is a law of physics that
\\
a. derives from quantum mechanics (a mathematically consistent theory
of physics) and 
\\
b. has been verified experimentally to great precision\cite{bollinger}.
\\
Therefore, the security of quantum cryptography was thought to
rest on very solid foundations.
\\
Here I proposed
a practical attack procedure that 
cannot be described without a theory of quantum gravity 
even approximately in general.
It breaks BB84
if gravitational nonlinearities exist.
However, neither
\\
a. do we know
a consistent theory of quantum gravity 
\\
nor
\\ 
b. was the linearity of 
evolution in the presence of gravitational fields
checked with the precision that can be achieved with
state-of-the-art equipment.
\\
Therefore presently
the security of quantum cryptography against this attack
can be guaranteed neither by recourse to general principles
nor by evaluating results of sensitive experimental tests.
\\
The latter gap could be quickly closed: 
experimental attacks on BB84 could assure at least
the practical (if not theoretical) 
security of this protocol. 
Such a test receives additional justification as
an unconventional search for experimental clues to quantum gravity.
\\
A complete theoretical treatment of the fundamental security
of quantum cryptography 
will only be possible when the correct theory of quantum gravity
is found. This raises a considerable practical interest in 
the most fundamental subject of contemporary 
physics.
If the security of quantum cryptography can be proved in the absence of a full theory
of quantum gravity, perhaps for other protocols than BB84, is an important
question for further research\cite{bhk_prep}.

\section{Acknowledgements}
I sincerely thank Don Page for a very helpful extended correspondence on his 
seminal experiment\cite{page81}, that was the model for the attack suggested here and
Claus Kiefer for useful comments on a manuscript draft.
Discussions with Jonathan Barrett and Adrian Kent about the security of 
quantum cryptography in the presence of nonlinearities were crucial
to my understanding of this issue.
Two anonymous referees helped to improve the manuscript with critical
comments.


\begin{thebibliography}{00}
\bibitem{qc} N. Gisin, G. Ribordy, W. Tittel, H. Zbinden 
{\it Quantum Cryptography}, Rev. Mod. Phys. {\bf 74}, 145-195 (2002)
\bibitem{lo_talk} H. Lo, abstract of talk:
{\it From quantum cheating to quantum security}, 
April 8, 2004, Los Alamos National Laboratory;
{\rm ``... quantum cryptography can come to the rescue 
by allowing perfectly secure communication guaranteed by the 
laws of physics.''}
\bibitem{christandl04} M. Christandl, R. Renner, A. Ekert,
{\it A Generic Security Proof for Quantum Key Distribution},
quant-ph/0402131 (2004); {\rm ``Quantum-key distribution provides perfect security because,
unlike its classical counterpart, it relies on the laws of 
physics rather than on ensuring that successful eavesdropping would require
excessive computational effort.''}
\bibitem{goldreich} O. Goldreich, S. Goldwasser,
{\it On the possibility of basing Cryptography on the 
assumption that P $\neq$ NP}, Cryptology ePrint Archive 1998/005 (1998)´
\bibitem{lo} H. Lo, H.F. Chau, {\it Unconditional Security Of Quantum Key 
Distribution Over Arbitrarily Long Distances},
Science {\bf 283},2050-2056 (1999)
\bibitem{mayers} D. Mayers, {\it Unconditional security in Quantum Cryptography},
JACM {\bf 48},351-406 (2001)
\bibitem{einstein} A. Einstein, {\it Grundz\"uge der Relativit\"atstheorie}, (Vieweg 
\& Sohn, Braunschweig,1956)
\bibitem{qg} C. Kiefer, {\it Quantum Gravity}, (Clarendon Press, Oxford, 2004)
\bibitem{page81}
D. Page, C. Geilker, {\it Indirect Evidence for Quantum Gravity}, 
Phys. Rev. Lett. {\bf 47},979-982 (1981) 
\bibitem{everett}
H. Everett III, {\it Relative State Formulation of Quantum Mechanics}, 
Rev. Mod. Phys. {\bf 29},454-462 (1957)
\bibitem{zeh70} H.D.Zeh,  
{\it On the interpretation of measurements in quantum theory},
Found. Phys. {\bf 1},69-76 (1970)
\bibitem{deutschb} D.Deutsch, {\it The Fabric of Reality: 
The Science of Parallel Universes - And Its Implications}, (Penguin, London, 1997).
\bibitem{preskill229} J. Preskill, Lecture Notes for Physics 229: {\it Quantum Information
and Computation},
Caltech, September 1998
\bibitem{dewitt} B.S. DeWitt, {\it God's Rays}, Physics Today,{\bf 58,1}, 32 (2005)
\bibitem{privQICL} priv. comm. with participants of the 
workshop ``Quantum information, computing
and logic'', July 2005, (Perimeter Institute, Waterloo)
\bibitem{wang}
C.H.-T. Wang, {\it Nonlinear quantum gravity on the constant mean curvature
foliation}, Class. Quantum Grav. {\bf 22},33-45 (2005)
\bibitem{cavaglia} M. Cavaglia, {\it Black hole and brane production in TeV gravity:
a review}, Int. J. Mod. Phys. A {\bf 18},1843-1882 (2003)
\bibitem{gisin} N.Gisin,
{\it Stochastic quantum dynamics and relativity}, Helv. Phys. Acta {\bf 62},
363-371 (1989)
\bibitem{kent_mixed} A. Kent,  {\it Nonlinearity without Superluminality},
Phys. Rev. A{\bf 72},012108 (2005)
\bibitem{polchinski}
J. Polchinski, {\it Weinberg's nonlinear quantum mechanics and the 
Einstein-Podolsky-Rosen paradox}, 
Phys. Rev. Lett. {\bf 66},397-400 (1991)
\bibitem{barrett} J.Barrett, A.Kent, L.Hardy, 
{\it No Signaling and Quantum Key Distribution},
Phys. Rev. Lett. {\bf 95}, 010503 (2005)
\bibitem{bollinger}J.J. Bollinger, D.J. Heinzen, W.M. Itano, 
S.L. Gilbert, and D.J. Wineland, {\it Test of the linearity of quantum 
mechanics by rf spectroscopy of the $^9$Be$^{+}$ ground state},
Phys. Rev. Lett. {\bf 63},1031-1034 (1989)
\bibitem{bhk_prep} J. Barrett, A. Kent, R. Plaga, under preparation
\end{thebibliography}
\end{document}